\documentclass[aps,pra,twocolumn,superscriptaddress]{revtex4}

\usepackage{makeidx}
\usepackage{amsfonts}
\usepackage{amssymb}
\usepackage{enumerate}
\usepackage{latexsym}
\usepackage{bbm}
\usepackage{graphicx}
\usepackage{graphics}
\usepackage{float}

\newcommand{\beq}{\begin{equation}}
\newcommand{\eeq}{\end{equation}}
\newcommand{\beqa}{\begin{eqnarray}}
\newcommand{\eeqa}{\end{eqnarray}}

\newcommand{\ket}[1]{\left\vert #1 \right\rangle}

\begin{document}

\title{Arbitrarily High Super-Resolving Phase Measurements at Telecommunication Wavelengths}

\author{Christian Kothe}
\affiliation{School of Information and Communication Technology, Royal Institute of Technology (KTH), Electrum 229, SE-164 40 Kista, Sweden}
\affiliation{Department of Physics, Stockholm University,
SE-109 61 Stockholm, Sweden}
\author{Gunnar Bj\"{o}rk}
\affiliation{School of Information and Communication Technology, Royal Institute of Technology (KTH), Electrum 229, SE-164 40 Kista, Sweden}
\author{Mohamed Bourennane}
\affiliation{Department of Physics, Stockholm University,
SE-109 61 Stockholm, Sweden}

\date{\today}

\begin{abstract}

We present two experiments that achieve phase super-resolution at
telecommunication wavelengths. One of the experiments is realized in
the space domain and the other in the time domain. Both experiments
show high visibilities and are performed with standard lasers and
single-photon detectors. The first experiment uses six-photon
coincidences, whereas the latter needs no coincidence measurements,
is easy to perform, and achieves, in principle, arbitrarily high
phase super-resolution. Here, we demonstrate a 30-fold increase of
the resolution. We stress that neither entanglement nor joint
detection is needed in these experiments, demonstrating that neither
is necessary to achieve phase super-resolution.

\end{abstract}

\pacs{42.50.Dv, 03.67.-a, 42.50.St}

\maketitle

\section{Introduction}

Interference plays a crucial role in many physical measurements,
such as detection of gravitational waves \cite{BW, HB, UK},
metrology \cite{WZ, GP}, interferometry and atomic spectroscopy
\cite{Do, CG, BI}, imaging \cite{Pl}, and lithography \cite{BK}.
Improvements of these schemes can be achieved with the help of {\it
phase super-resolution}, where $n$ oscillations (fringes) appear in
the interference pattern over a range which usually would have given
only one oscillation \cite{WP,Mitchell}. In a similar vein, {\it
phase super-sensitivity} decreases the phase uncertainty in such
experiments, so that the measurement sensitivity would surpass the
classical limit, i.e., beating the standard quantum limit
\cite{RP,Okamoto}.

It was believed that entangled states are needed to achieve phase
super-resolution \cite{GL}. One such state is a path-entangled
number-state, the so called NOON-state \cite{Noon} \beq\label{eq1}
\ket{\Psi(0)}=\frac{1}{\sqrt{2}}\left(\ket{N}\ket{0}+\ket{0}\ket{N}\right),
\eeq where $N$ denotes the number of particles (most often photons).
This state is a superposition of $N$ particles in one path and no
particles in the other path, and vice versa. The production of such
states at satisfying rates gets extremely difficult already for
small $N$. So far, only experiments with $N$ up to 4 have been
reported \cite{WP, NO}. Recently, however, schemes were proposed and
shown, where phase super-resolution could be achieved by
unentangled, coherent light \cite{RP}. In that paper an experiment
was reported, where $n=N=6$ oscillations occur over the period of
one classical oscillation. For their experiment, Resch et. al. used
6-photon coincidence, but got rather limited counting rates ($\leq
27$ counts/10 s) and moderate visibility (50-90 \%). Other
experiments with unentangled states and phase super-resolution with
$n>4$ have been reported \cite{HB2}. All of those experiments used
light at wavelengths around $800$ nm.

In this letter, we report on two super-resolution experiments. One
is in the {\it space domain} like all the experiments hitherto
reported, but using an innovative approach resulting in higher
counting rates and very high visibility. The other experiment is
performed in the {\it time domain} and shows very promising counting
rates and visibility. Both of the experiments could in principle be
scaled to high numbers of $n>100$. Whereas the first experiment
requires more components for higher numbers of $n$, the second
experiment only requires a longer measurement time. Our space- and
time-domain experiments were performed up to $n=6$ and $n=30$,
respectively. Furthermore, these experiments are, to the best of our
knowledge, the first ones performed at telecommunication wavelengths
which allows efficient transmission of the photons over long
distance via optical fiber and thereby allow more opportunities for
applications of phase super-resolution.

\section{Phase super resolution - ``dequantified''}

We briefly introduce the theory of phase super-resolution.

Consider the state given in Eq. (\ref{eq1}). Imposing a relative
phase-shift of $\phi$ between the two modes transforms the state
into \beq \label{eq2}
\ket{\Psi(\phi)}=\frac{1}{2}\left(\ket{N}\ket{0}+e^{i
N\phi}\ket{0}\ket{N}\right), \eeq since energy (difference) is the
generator of (relative) phase. The phase $N \phi$ will therefore
grow linearly with the number of particles $N$. Treating the two
modes for example as spatial modes and combining them via a 50/50
beamsplitter would result in a detection probability $P\propto
1\pm\cos\left(N\phi\right)$ in the two outputs of the beamsplitter
when measuring $N$-fold coincidence. $P$ will exhibit phase
super-resolution, since it oscillates $n=N$ times when $\phi$ varies
from 0 to $2\pi$. The same effect, however, can also be reached
without entanglement by a so-called time-reversal measurement
\cite{RP}. In the cited paper the effect is explained by using the
inherent time-reversal symmetry of quantum mechanics and measurement
of entanglement. A different view of the experiment is the
following, based on the mathematical relation
$\sin(2\phi)=2\sin(\phi)\cos(\phi)$, or, more generally, \beqa
\nonumber
\frac{\sin\left(n\phi\right)}{2^{n-1}}& = & \sin\left(\phi\right)\sin\left(\phi+\frac{\pi}{n}\right)\sin\left(\phi+\frac{2\pi}{n}\right) \cdots\\
\label{eq3} &&\cdots \sin\left(\phi+\frac{(n-1)\pi}{n}\right).
\label{Eq: sin}\eeqa A phase super-resolving measurement can hence
be implemented as a multiplication (e.g., coincidence detection) of
$n$ ordinary phase measurements, each shifted by $k \pi/n$, where
$k=0,1, \ldots , n-1$.

A coherent state can be split into several modes, where each of
these will be a coherent state \cite{Glauber}. Since the ensuing
multi-mode state is separable, there are no quantum correlations
between the states.
\begin{figure}
\center
\includegraphics[angle=0, scale=.65]{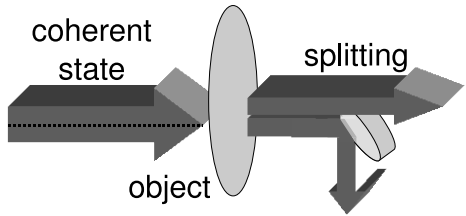}
\caption{A coherent state split spatially into two by a mirror.}
\label{Fig: splitting}
\end{figure}
E.g., a coherent-state in an optical beam can be split in two halves
by a mirror, as in Fig. \ref{Fig: splitting}. If the two ``half
beams'' come from the same or from different, identical sources
makes no difference. It also does not matter if the beam is split
after the interaction by a totally reflecting mirror inserted
halfway into the beam, or by a semitransparent beam splitter,
insofar the measured object characteristic does not vary over the
width of the beam. Hence, we may as well split the beam already
before the object, as in Fig. \ref{setup_space}. It also does not
matter if the coherent state mode is split spatially, temporally or
in frequency insofar the object does not vary over the corresponding
space-, time-, or frequency range. In previous experiments the state
has been split spatially, by beam splitters, after the interaction.
We find it simpler to split the state before the interaction, or in
time, and this and the observation delineated in Eq. (\ref{Eq: sin})
are the basic ingredients of our experiments and the interpretation
thereof.

\section{experiment - space domain}

\begin{figure}
\center
\includegraphics[angle=0, scale=.45]{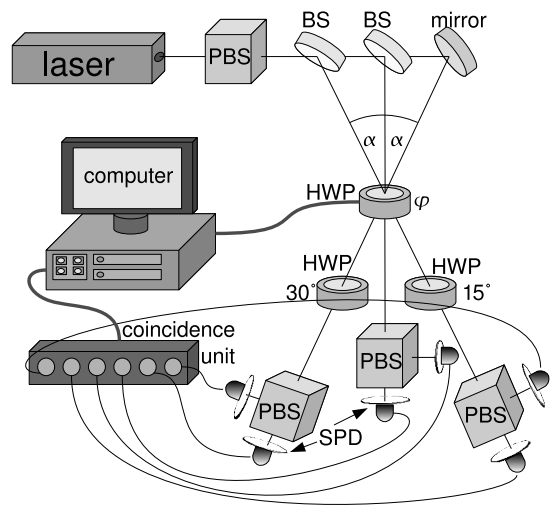}
\caption{Setup to measure 6-fold phase super-resolution in the space domain. See the text for further details.}
\label{setup_space}
\end{figure}
\begin{figure}
\center
\includegraphics[angle=0, scale=.8]{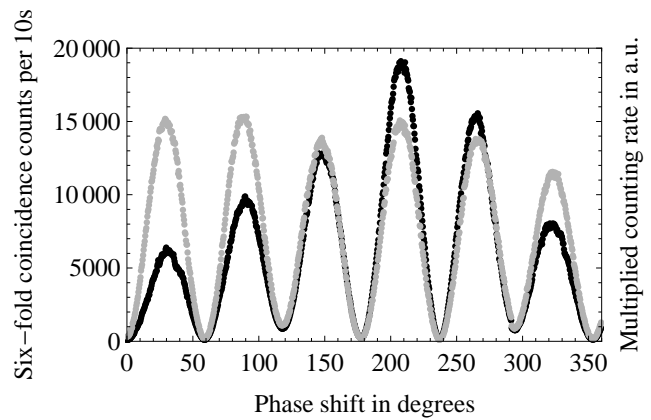}
\caption{Phase super-resolution in the space-domain by measuring 6-fold coincidence (grey dots) and by multiplying single-photon detections (black dots). The statistical errors are smaller than the dot size.}
\label{results_space}
\end{figure}
In our experiment, to measure phase super-resolution in the space
domain we built the setup illustrated in Fig. \ref{setup_space}.
Except for the fact that we split the coherent state before the
half-wave plate (HWP), the setup is essentially identical to that in
Ref. \cite{RP}. The coherent state is generated by a $\lambda =
1550$ nm laser, which is pulsed at a rate of $2.5 {\rm\ MHz}$. Every
pulse has a duration of around $500 {\rm\ ps}$ and the average power
of the laser is $1 {\rm\ mW}$, which, directly after the laser, is
attenuated by $50-55 {\rm\ dB}$ (not shown in the figure) to avoid
``overexposure'' of our single-photon detectors (SPD). A polarised
beam-splitter (PBS) in front of the laser assures that the light is
horizontally (H) polarised. Two 50/50 beam-splitters (BS) divide the
beam into three paths, which are superimposed on a HWP. The HWP can
be rotated to any angle $\varphi$ by a motor. One has to assure that
the angle $\alpha$ between the beams is small so that the
pathlengths of the beams in the HWP (the object) are essentially the
same. In our case, we use an angle of $\alpha=2.9^{\circ}$. Since
$\varphi=45^{\circ}$ turns the polarisation from H to vertical (V)
one can see one oscillation by turning the HWP by $90^{\circ}$
(corresponding to a relative-phase shift of $360^{\circ}$ in our
figures) and detect, for example, only H-polarised photons. Two of
the beams pass another HWP rotated by the angles $15^{\circ}$ and
$30^{\circ}$, respectively, to rotate the polarization further in
such a way that they fullfill the relation in Eq. (\ref{eq3}) after
detection. Each beam then pass another PBS, carefully oriented so
that the H-polarised light goes into one arm and the V-polarised
light into the other arm. These 6 beams are then coupled into
single-mode fibres which lead to SPDs. The SPDs are gated and open
only for 1~ns when a coherent-state pulse is coming, and the outputs
of the SPDs are led to a six-channel coincidence counter, and
subsequently to a computer where the coincidences are registered and
stored.

The results can be seen in Fig. \ref{results_space}. Every SPD
detected one oscillation while turning the HWP by $90^{\circ}$
(corresponding to a phase shift of 360 degrees in the figure), but
from one SPD to the next the peak of the oscillation was shifted by
$60$ degrees. Multiplying the stored, individual counts from all the
six SPDs in the computer gives the black dots in the figure, where
the scale is in arbitrary units. In the case of the 6-fold
coincidences (grey dots) the scale to the left applies. One can
clearly distinguish 6 peaks as expected, so a sixfold increase of
the phase resolution is achieved. At every angle we were measuring
for 10 seconds and got counting rates of more than 1500 counts/s for
the 6-fold coincidences and about $10^6$ counts/s for the individual
SPDs. The visibility is high, between 98.6~\% and 96.7~\% in the
case of the multiplication of the single detections and between
98.6~\% and 97.0~\% in the case of the coincidence detection. For
calculating the visibility we took the lowest of the to minima
around a peak. The reason that two of the minima are higher than the
other ones is due to imperfections in one of the PBSs, which does
not perfectly split the beam into H- and V-polarised components.
Having better PBSs would probably result in a more uniform
visibility $>$98~\%. Another improvement would be to replace the
first 50/50 BS with a 33/66 BS so that all the beams have the same
intensity. This would neither change the visibility nor the
resolution, but would increase the coincidence rate. In principle,
the setup can be extended to detect phase super-resolution with a
factor $n>6$ by adding more arms. The drawbacks are that this
requires more components and that the coincidence counting-rate gets
exponentially lower so that one would have to increase the
measurement time correspondingly.

\section{experiment - time domain}

\begin{figure}
\center
\includegraphics[angle=0, scale=.45]{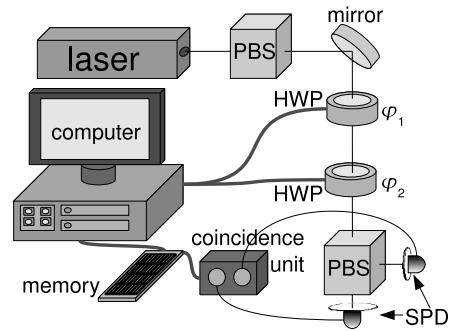}
\caption{Experimental setup to measure phase super-resolution in the time domain. See the text for further details.}
\label{setup_time}
\end{figure}
\begin{figure}
\center
\includegraphics[angle=0, scale=.8]{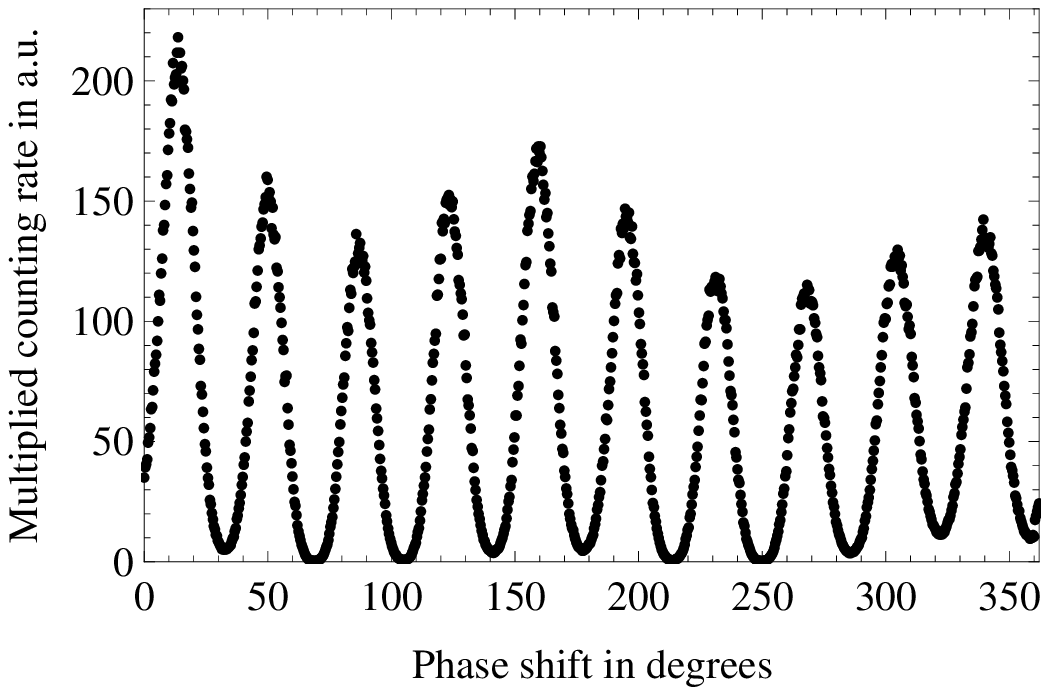}
\includegraphics[angle=0, scale=.8]{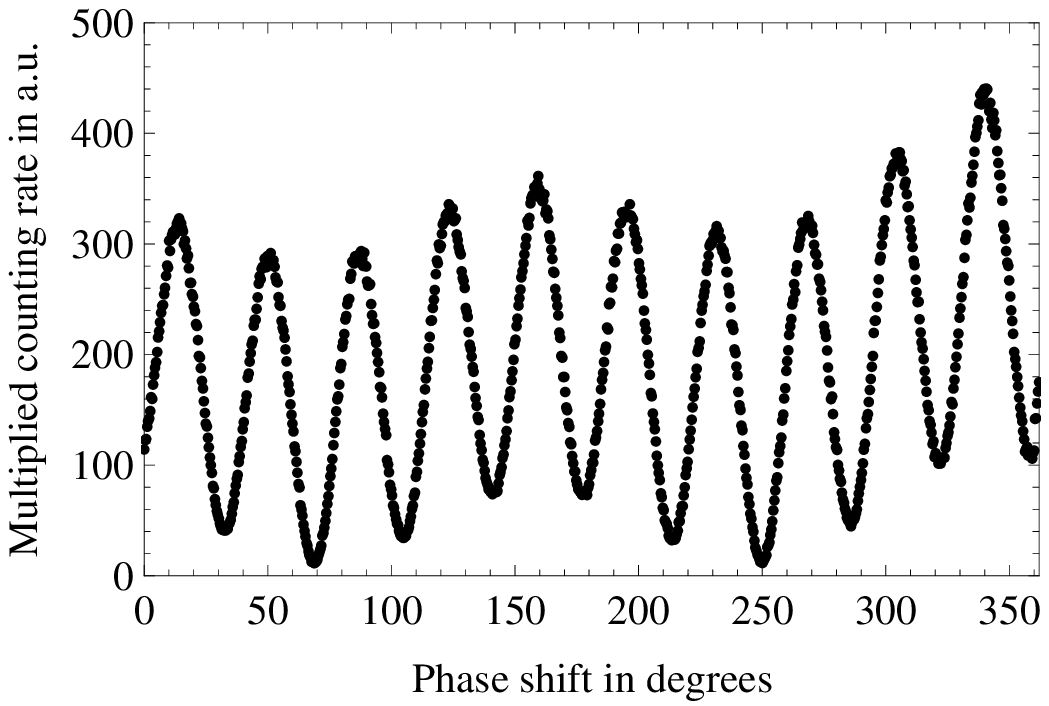}
\caption{Phase super-resolution with $n=10$ in the time domain. The data was taken with the help of two SPDs and a coincidence unit (upper panel) and with the help of only one SPD (lower panel).}
\label{results_time}
\end{figure}
\begin{figure}
\center
\includegraphics[angle=0, scale=.8]{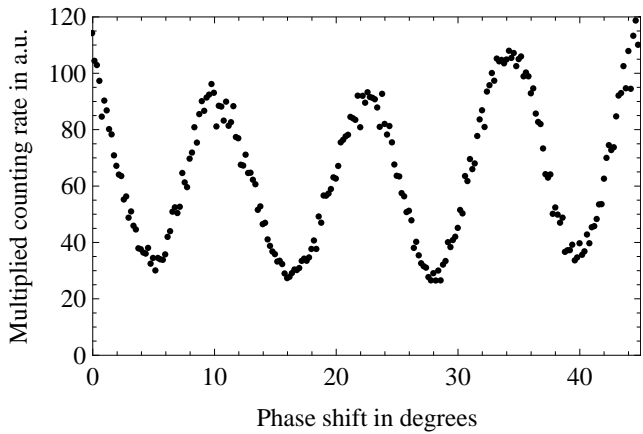}
\caption{Phase super-resolution in the time domain with a resolution enhancement by a factor of $n=30$.}
\label{results_time2}
\end{figure}

To obtain phase super-resolution with $n>6$ it therefore is more
practical to do the measurement in the time domain. To this end we
built the setup in Fig.~\ref{setup_time}. The parameters of the
laser are the same as in the setup in the space domain, except a
slightly higher attenuation. Instead of having several physical arms
in the setup we used a computer to store the data sequentially as to
have $n$ ``arms after each other'' in time. In this setup we counted
and stored the clicks in each SPD as a function of $\varphi_1$,
turned the second HWP by an angle $\Delta\varphi_2=90^\circ/n$,
repeated the process $n$ times and numerically multiplied the
results according to equation (\ref{eq3}). For $n=6$ we set the
second HWP at the angles $\varphi_2=0^\circ, 15^\circ {\rm\ and\ }
30^\circ$ using both the SPDs, or at $\varphi_2=0^\circ, 15^\circ,
30^\circ, \ldots , 75^\circ$ when using only one SPD.

The results of our measurements can be seen in
Fig.~\ref{results_time}. We were rotating the first HWP by small
increments over the range $0 \leq \varphi_1\leq 90^\circ$ to
introduce a differential phase-shift. Then we change the angle of
the second HWP after each scan of $\varphi_1$ by
$\Delta\varphi_2=9^\circ$ to get phase super-resolution by a factor
of $n=10$. At each  combination of angles we measured for one
second. The upper panel of Fig.~\ref{results_time} shows the results
when we used both SPDs followed by a coincidence counter, and the
lower panel shows the results when we used only one SPD, requiring
twice as many settings of the angle $\varphi_2$. As one can see, the
first method gives higher visibility, 99.6~\%, and takes only half
the time, and is therefore more efficient. On the other hand does
the second method need neither the second SPD nor the coincidence
detector in Fig.~\ref{setup_time}. This clearly shows that phase
super-resolution can be achieved with neither entanglement nor joint
detection.

Further results can be seen in Fig.~\ref{results_time2}, where we
achieved phase super-resolution by a factor of $n=30$. (Note that
the $x$-axis in this case only spans the range $0^\circ$ to
$45^\circ$.) In this case too, every combination of angles
$\varphi_1,\varphi_2$ was measured for one second. The reason that
the visibility is degraded is most certainly due to the insufficient
stability of our laser's intensity over long times. Since higher
values of $n$ in our experiments require smaller measurement
increments of $\varphi_1$ to resolve one oscillation period, we had
to measure for a long time to get phase super-resolution with $n=30$
over the whole range of $\varphi_1$ between 0 and 360 degrees. (1800
settings of $\varphi_1$, 30 settings of $\varphi_2$ and 1~s
measurement time per setting plus time for rotation gives around 20
hours.) Under practical conditions, however, this should not pose
any substantial problem, since one usually is not interested in
scanning the whole range between 0 and 360 degrees with phase
super-resolution, but rather make some rough scans to start with,
and then limit oneself to a narrow range of angles $\varphi_1$ or
phases. Furthermore, in most cases, measurement times of less than
one second per setting should  give sufficient visibility.

An advantage of the time-domain method is that one can get phase
super-resolution with high factors $n$. A technical limitation of
our setup is the stepsize resolution $\Delta\varphi_1$ and
$\Delta\varphi_2$ in turning the HWPs. Stepsizes of
$\Delta\varphi=0.1^\circ$ posed no problem in our setup,
theoretically the rotator specification allows a resolution of
around one minute of arc. If $m$ denotes the number of points in
each fringe this gives the possibility to achieve
$n=90/(m\Delta\varphi)$, for example $m=3$ and one minute of arc as
the stepsize of the rotator would allow $n=1800$. A more fundamental
problem is the measurement time. The object under investigation
should not undergo any changes in phase during the measurement time.
Additionally, the intensity of the laser should be stable during the
whole measurement time because the SPDs have no way to tell whether
a change in the counting rate is due to a phase change or on due to
an intensity variation of the laser. Since our laser was not
sufficiently stable over long times we could not use the full
advantage of the small stepsizes achievable by our rotators, but
``only'' achieved $n=30$.

We stress that our experiments did not manifest any phase
super-sensitivity since we used coherent states, subject to shot
noise. For phase super-sensitivity it seems that entanglement is
required \cite{NO}.

\section{Interpretation}

Phase super-resolution turns out to be quite a simple and mostly
classical phenomenon. We have shown that neither entanglement nor
joint detection is needed to achieve it (note that the coincidence
unit and the second detector are not necessary in
Fig.~\ref{setup_time}, as explained in the previous section). There
is neither a need for time-reversal symmetry as introduced in
\cite{RP} nor, e.g., measurement induced, post-selection
entanglement. Basically everything one has to do is to shift
sine-functions as in Eq. (\ref{Eq: sin}) and multiplying them. This
task can be done with quite ordinary standard optical components and
multiplication, either by coincidence detection or by numerical
multiplication of the acquired data in a computer.

In principle these findings were known since Glauber's pioneering
work on quantum optics \cite{Glauber}. The correlation functions
$\left|g^{(n)}\left(x_1\cdots x_{2n}\right)\right|$ of any order
$n$, where $x$ denotes the time- and space-coordinates are symmetric
in time and space. This, together with Glauber's finding that
$n$-fold delayed coincidences which are ``detected by ideal photon
counters reduce to a product of the detection rates of the
individual counter'' leads to the fact that our experiments in time-
and space-domain give the same kind of physics as all the other
hitherto reported experiments, yet with much better results. We have
capitalized on this knowledge for the design and explanation of our
setup, that outperforms the experiments shown so far, and we hope
that it will clarify the requirements for achieving super-resolving
phase measurements.

\section{Conclusion}

In summary, we showed that super-resolving phase measurements with
simultaneously high $n$ (denoting the relative increase in the
number of fringes) and high visibility can be achieved. Whereas
there are clear limitations to the increase of $n$ in the space
domain, it is less demanding to reach high values of $n$ in the time
domain. We showed the principle up to $n=30$ and explained how one
could reach higher values without requiring any extra components by
just using a stabilized laser and high quality optics. We
furthermore performed the experiment at telecommunications
wavelengths, enabling possible remote applications since the
phase-shifting (including the polarization analyzers) and the
detectors, coincidence unit and control computer can be stationed at
different, remote locations.

\section{Acknowledgements}

CK acknowledges support from the German National Academic Foundation. The work was supported by the Swedish Research Council (VR) and by the Knut and Alice Wallenberg Foundation (KAW). We thank E. Amselem, M. R\aa dmark and J. Ahrens for discussions.

\end{document}